\documentclass[
 aip,
 amsmath,amssymb,
 reprint,%
]{revtex4-1}

\usepackage{upgreek}
\usepackage{graphicx}% Include figure files
\usepackage{dcolumn}% Align table columns on decimal point
\usepackage{bm}% bold math
\usepackage{braket}
\DeclareMathSymbol{\shortminus}{\mathbin}{AMSa}{"39}
\usepackage[utf8]{inputenc}
\usepackage[T1]{fontenc}
\usepackage{mathptmx}

\makeatletter
\DeclareRobustCommand*\uell{\mathpalette\@uell\relax}
\newcommand*\@uell[2]{
  \setbox0=\hbox{$#1\ell$}
  \setbox1=\hbox{\rotatebox{10}{$#1\ell$}}
  \dimen0=\wd0 \advance\dimen0 by -\wd1 \divide\dimen0 by 2
  \mathord{\lower 0.1ex \hbox{\kern\dimen0\unhbox1\kern\dimen0}}
}

\begin{document}

\preprint{AIP/123-QED}

\title{Wavelength transduction from a 3D microwave cavity to telecom using piezoelectric optomechanical crystals}

\author{H. Ramp}
\email{ramp@ualberta.ca}
\affiliation{Department of Physics, University of Alberta, Edmonton, Alberta T6G 2E9, Canada}
 
\author{T. J. Clark}%
\affiliation{Department of Physics, University of Alberta, Edmonton, Alberta T6G 2E9, Canada}

\author{B. D. Hauer}%
\affiliation{Department of Physics, University of Alberta, Edmonton, Alberta T6G 2E9, Canada}

\author{C. Doolin}%
\affiliation{Department of Physics, University of Alberta, Edmonton, Alberta T6G 2E9, Canada}

\author{K. C. Balram}
\affiliation{Center for Nanoscale Science and Technology, National Institute for Standards and Technology, Gaithersburg, Maryland 20878, USA}

\author{K. Srinivasan}
\affiliation{Center for Nanoscale Science and Technology, National Institute for Standards and Technology, Gaithersburg, Maryland 20878, USA}

\author{J. P. Davis}
\email{jdavis@ualberta.ca}
\affiliation{Department of Physics, University of Alberta, Edmonton, Alberta T6G 2E9, Canada}

%\date{\today

\begin{abstract}
Microwave-to-optical transduction has received a great deal of interest from the cavity optomechanics community as a landmark application for electro-optomechanical systems. In this Letter, we demonstrate a transducer that combines high-frequency mechanical motion and a microwave cavity. The system consists of a 3D microwave cavity and a gallium arsenide optomechanical crystal, which has been placed in the microwave electric field maximum. This allows the microwave cavity to actuate the gigahertz-frequency mechanical breathing mode in the optomechanical crystal through the piezoelectric effect, which is then read out using a telecom optical mode. The gallium arsenide optomechanical crystal is a good candidate for low-noise microwave-to-telecom transduction, as it has been previously cooled to the mechanical ground state in a dilution refrigerator. Moreover, the 3D microwave cavity architecture can naturally be extended to couple to superconducting qubits and to create hybrid quantum systems.
\end{abstract}

\maketitle

The field of quantum information has evolved along two dominant paths: microwave quantum technology has produced advanced quantum computers comprised of superconducting qubits that operate at frequencies between one to ten gigahertz. \cite{Arute2019} Meanwhile, optical quantum technology, at visible and telecom wavelengths, has enabled long-lived quantum memories, \cite{Saglamyurek2019} entanglement-based telescopes with sensitivity beyond classical limits, \cite{Gottesman2012} and high-precision atomic clocks. \cite{Komar2014} Linking these technologies through a high-efficiency, low-noise, coherent microwave-to-telecom transducer would open a new frontier of hybrid quantum technology.\cite{Kurizki2015} Prominent examples of hybrid technology include quantum radar\cite{Barzanjeh2015} and microwave quantum repeaters,\cite{Kumar2019} which would allow for distributed quantum computing\cite{DiVincenzo} and the creation of a quantum internet.\cite{Kimble2008,Lauk2019}

The development of a quantum microwave-to-telecom transducer has been pursued in a wide breadth of physical systems including cold atoms, \cite{Petrosyan2019} nonlinear electro-optic materials, \cite{Rueda2016,Javerzac-Galy2016, Fan2018} and mechanical resonators\cite{Pitanti2015,Bagci2014, Higginbotham2018,Andrews2014,Forsch2019,Balram2016,Bochmann2013, Vainsencher2016, Jiang2019a, Jiang2019b} with frequencies ranging from kilohertz to tens of gigahertz. Of these techniques, mechanical motion in the form of quantized phonons provides a promising avenue for quantum microwave-to-telecom transduction due to the ability to facilitate strong interactions between phonons and microwave \cite{Teufel2011} or telecom \cite{Groblacher2009} photons. Mechanically mediated microwave-to-telecom transducers can be classified into two groups according to their mechanical frequency. Mechanical modes with frequencies below $\approx1~\mathrm{GHz}$ use microwave resonators to bridge the energy differential between the qubit-emitted gigahertz photons and the megahertz mechanical phonons.\cite{Reed2017} Low frequency mechanical modes, such as the drum modes of a membrane, are advantageous in that they can be directly incorporated into a microwave resonator by depositing an electrode on the membrane to create a mechanically compliant capacitor in an LC resonator.\cite{Bagci2014, Higginbotham2018,Andrews2014} However, these low-frequency mechanical modes are also inherently problematic due to high thermal phonon occupations at dilution refrigerator temperatures, which add noise to transduced signals. \cite{Andrews2014} Conversely, mechanical modes with frequencies exceeding $\approx1~\mathrm{GHz}$ can be cooled to the ground state using a dilution refrigerator, \cite{Ramp2019, Forsch2019} but typically couple directly to on-chip microwave waveguides. \cite{Forsch2019,Balram2016,Bochmann2013, Vainsencher2016, Jiang2019a, Jiang2019b}

In this Letter, we take the natural step to combine these architectures and implement a high-frequency mechanically-mediated microwave-to-telecom transducer coupled to a microwave resonator. The transducer, which is operated at room temperature in this experiment, consists of a gallium arsenide optomechanical crystal placed on the central pedestal of an aluminium re-entrant 3D microwave cavity. Gallium arsenide is a promising material for microwave-to-telecom transducers due to its piezoelectric \cite{Masmanidis2007} and photoelastic\cite{Balram2014} properties, which respectively allow mechanical motion to be coupled to the electric field of the 3D microwave cavity mode and improves coupling to the telecom optical mode of the optomechanical crystal. The 3D microwave cavity architecture\cite{LeFloch2013} is a versatile platform for quantum systems---in addition to providing a coupling mechanism for qubits,\cite{Paik2011} they are also used to couple microwave photons to magnons in yttrium-iron-garnat spheres,\cite{Goryachev2014,Tabuchi2014,Zhang2014,Kostylev2016} nitrogen-vacancy centers in diamond,\cite{Angerer2016} rubidium atoms,\cite{Tretiakov2020} and Rydberg atoms.\cite{Stammeier2017}

\begin{figure*}[t]%[width=1.95\columnwidth]
\includegraphics{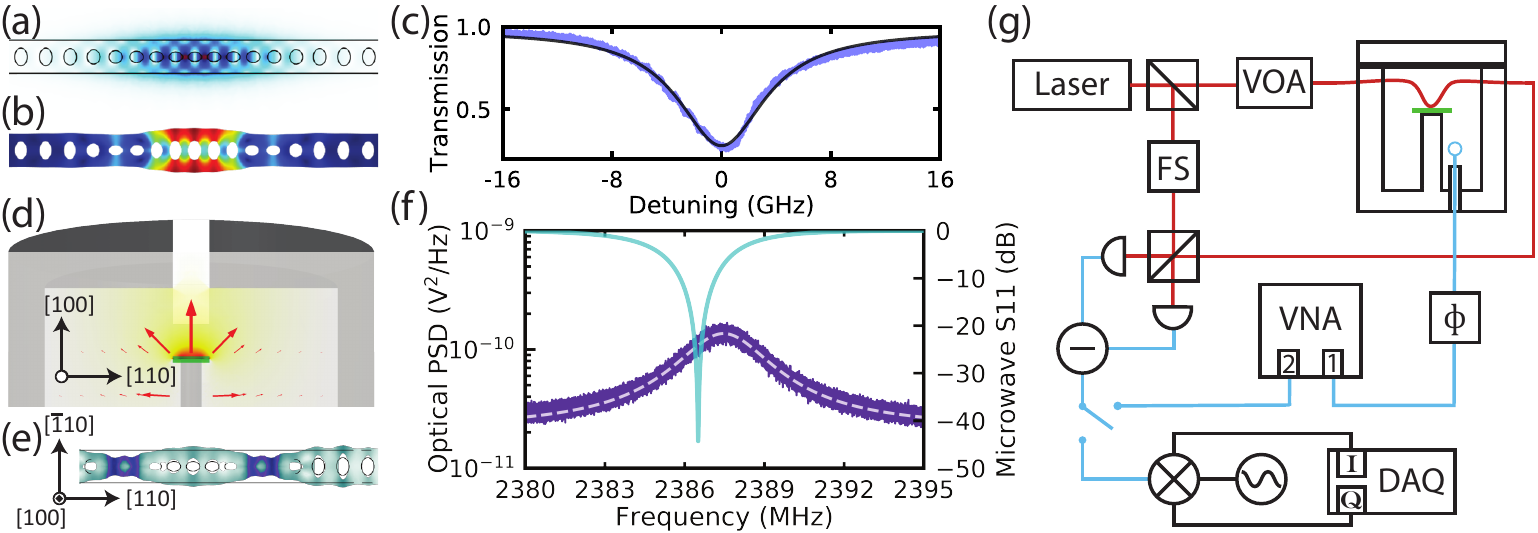}
\caption{\label{fig:1} (a)~Simulated telecom electric field and 
(b)~mechanical displacement modes of the optomechanical crystal. 
(c)~Transmission of the telecom resonance, centered at $1543~\mathrm{nm}$. 
(d)~Simulation of the electric field in the 3D microwave cavity mode with arrows showing electric field directivity and relative amplitude. 
(e)~Simulated piezoelectric response to a resonant microwave drive, where a time-varying uniform electric field is applied along [100].
(f)~Purple trace (left axis): power spectral density (PSD) of the thermomechanical motion, measured by homodyne detection of the telecom mode and fit (white).
Light blue trace (right axis): microwave reflection measurement of the 3D cavity.
(g)~Measurement setup: the microwave system in blue shows port 1 of a vector network analyser (VNA) driving the microwave cavity using a loop coupler. The VNA signal phase is controlled externally using a phase modulator ($\upphi$). The balanced laser homodyne system in red follows two paths: the measurement arm, with telecom optical power set by a variable optical attenuator (VOA) before coupling to the optomechanical crystal (green) in the microwave cavity, and the local oscillator, with a fiber stretcher~(FS) for path-length matching and optical phase control. The paths recombine at a beam splitter and are then detected on a balanced photodiode. The photodiode output can either be measured on VNA port 2 or downmixed into low frequency in-phase and quadrature components measured on a separate data acquisition (DAQ) system.}
\end{figure*}

The optomechanical crystal geometry, which measures $600~\mathrm{nm}$ wide, $220~\mathrm{nm}$ thick, and $13~\upmu \mathrm{m}$ long, is engineered to create a band-gap that supports a telecom optical mode\cite{Balram2014} at $1543~\mathrm{nm}~(\omega_\mathrm{c}/2\pi \approx194.3~\mathrm{THz})$, Fig.~\ref{fig:1}(a). The telecom mode overlaps a high frequency mechanical breathing mode at $\omega_\mathrm{m}/2\pi \approx 2387.5~\mathrm{MHz}$, with damping rate $\Gamma_\mathrm{m}/2\pi \approx 2.90~\mathrm{MHz}$, Fig.~\ref{fig:1}(b). The laser transmission sweep of the telecom mode in Fig.~\ref{fig:1}(c) demonstrates a total decay rate $\kappa/2\pi \approx 6.6~\mathrm{GHz}$ and external decay rate $\kappa_\mathrm{e}/2\pi \approx 2.3~\mathrm{GHz}$.

The 3D microwave cavity, presented in Fig.~\ref{fig:1}(d), has an inner diameter $41~\mathrm{mm}$ and an interior height $35~\mathrm{mm}$. The re-entrant pedestal measures $2.8~\mathrm{mm}$ in diameter and stands $25~\mathrm{mm}$ tall, which leaves a $10~\mathrm{mm}$ gap between the cavity lid and the top of the re-entrant pedestal where the optomechanical crystal is placed. The cavity lid is split a by $5~\mathrm{mm}$ gap which allows a dimpled tapered fiber\cite{Hauer2014} to be lowered into the cavity for optical coupling. Despite the gap, the microwave mode electric field, shown using arrow vectors in Fig.~\ref{fig:1}(d), is predominantly directed along the cylinder axis, which corresponds to the $[100]$ crystal axis of the gallium arsenide chip. This design results in a microwave resonant frequency $\omega_{\upmu}/2\pi \approx 2386.5~\mathrm{MHz}$. A low-loss teflon cylinder \cite{Carvalho2019} surrounding the re-entrant pedestal allows for control over the microwave cavity decay rates by shaping the magnetic field, such that the total cavity decay rate $\kappa_\mathrm{\upmu}/2\pi \approx 4.07~\mathrm{MHz}$ is almost exactly double the external decay rate $\kappa_\mathrm{\upmu,e}/2\pi \approx 2.05~\mathrm{MHz}$. This condition, known as critical coupling, ensures that on-resonance there is near-zero ($-45~\mathrm{dB}$) reflection of microwave power from the cavity.\cite{Aspelmeyer2014}
    
Coupling between the microwave electric field and the mechanical motion of the optomechanical crystal is mediated by the $d_{36}$ coefficient of the piezoelectric tensor, which converts a transverse electric field oriented along the $[100]$ crystal axis into shear motion in the plane of the optomechanical crystal.\cite{Masmanidis2007} In Fig.~\ref{fig:1}(e), the displacement of the optomechanical crystal due to a resonant transverse electric field is simulated to demonstrate the similarity between the mechanical breathing mode and microwave driven motion. The spatial overlap between the mechanical mode and driven motion indicates that the piezoelectric interaction between the microwave transverse electric field and the mechanical breathing mode should result in well-coupled modes.  Although the microwave cavity electric field is predominantly transverse, small in-plane electric field components caused by imperfect electric field directivity are capable of driving high-order flexural and torsional modes in the optomechanical crystal, however these modes have poor overlap with the optical mode and are therefore not measured. 

In experiment, we find that the microwave mode is slightly detuned from the mechanical resonance such that $(\omega_\upmu - \omega_\mathrm{m})/2\pi \approx 1~\mathrm{MHz}$. Despite this small offset, Fig.~\ref{fig:1}(f) shows that the microwave mode frequency lies within the linewidth of the mechanical mode and vice versa, implying that a signal can be resonantly enhanced by both the microwave and mechanical modes. The measurement setup used in this experiment is presented in Fig.~\ref{fig:1}(g), which illustrates the homodyne laser system used to measure the optomechanical crystal and the microwave network used to drive the microwave cavity.

\begin{figure}[t]%.96
\includegraphics[width=.95\columnwidth]{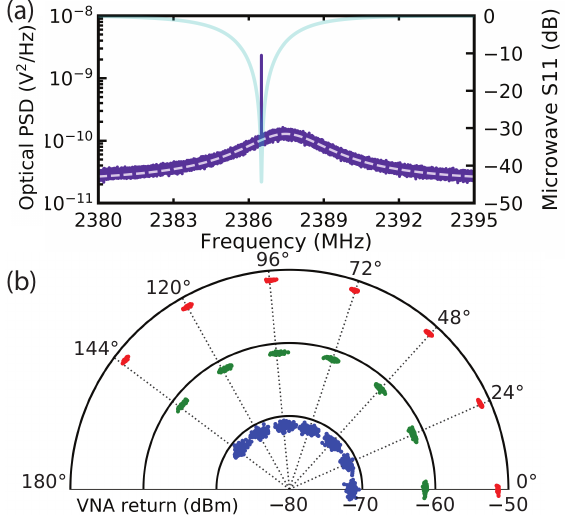}
\caption{\label{fig:2} (a) Microwave-to-telecom transduction. A $10~\mathrm{\upmu W}$ electrical signal at the microwave resonance frequency creates a sharp peak in the optical PSD.  (b) Phase coherence measurements of the transduced signal. The input signal phase (dotted lines) is swept in $24^\circ$ steps using the external modulator, for three sets of microwave input power: $200~\mathrm{\upmu W}$~(red), $20~\mathrm{\upmu W}$~(green), and $2~\mathrm{\upmu W}$~(blue).
}
\end{figure}

In Fig.~\ref{fig:2}(a), the mechanical motion of the optomechanical crystal is measured while the microwave cavity is driven by a $-20~\mathrm{dBm}$ electrical signal at frequency $\omega_\mathrm{s}$, which is set to the microwave resonance frequency. The electrical signal populates the microwave mode with photons, which are converted into actuated phonons in the mechanical mode through the piezoelectric effect. The telecom laser, at frequency $\omega_{\uell}$, causes the actuated phonons to be up-converted into telecom photons at frequency $\omega_{\uell} + \omega_\mathrm{s}$. The up-converted photons beat together with a local oscillator to produce a sharp peak in the homodyne measurement---the transduced microwave tone.

Coherent transduction is then demonstrated by using an external phase modulator to sweep the phase of the injected microwave signal, plotted in Fig.~\ref{fig:2}(b), from $0^{\circ}$ to $144^{\circ}$. The output of the balanced photodiode is returned to the second port of the VNA for an S21 measurement, where the returned signal mixes with the VNA output signal to determine relative phase. By manipulating both the phase and amplitude, we show that we have complete coherent control of the transduced signal.\cite{Balram2016} Classically, coherent control implies that the system can be used for information transduction techniques such as phase shift keying.\cite{Rudd2019}

\begin{figure}[b]
\includegraphics[width=.95\columnwidth]{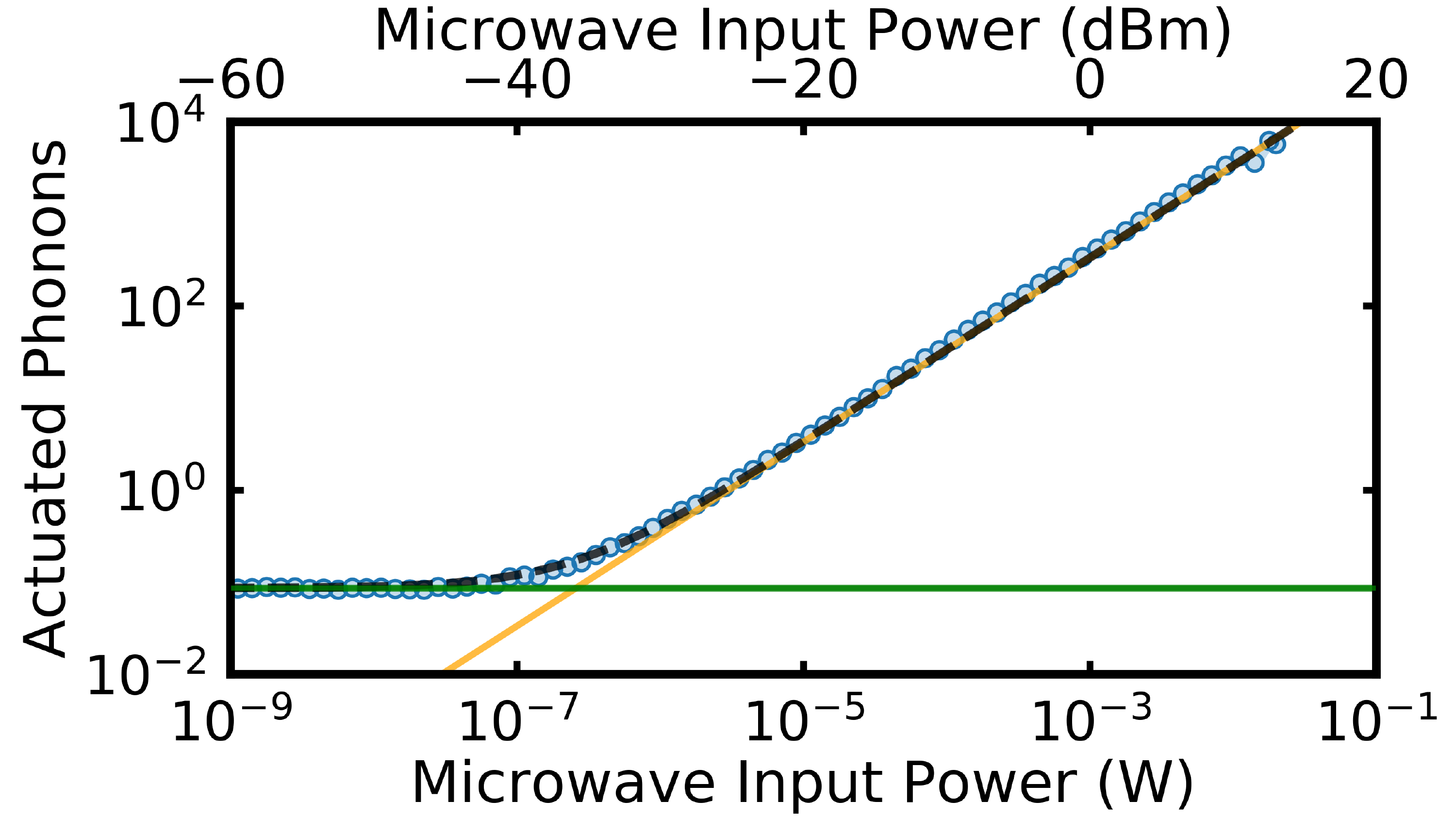}
\caption{\label{fig:3}The integrated area of the transduced peak, scaled using Eqn.~\eqref{eqn:PSDtophonons} to measure the actuated phonon number, measured as microwave power is stepped. The resulting trace is fit in two segments: input power above $10^{-5}~\mathrm{W}$ is fit to a line (orange) to determine the electromechanical coupling, and input power below $10^{-8}~\mathrm{W}$ is fit to a constant to determine the measurement noise floor, below which the transduced signal becomes unmeasurable due to thermal noise. The sum of these fits is presented as the black dashed curve. Error in the actuated phonon number is smaller than the marker size. The uncertainties of derived quantities are calculated from $1\sigma$ fit error.
}
\end{figure}

The number of actuated phonons $\bar{n}_\mathrm{s}$ in the mechanical breathing mode can be calculated using thermomechanical noise as a calibration metric.\cite{Balram2016}  The ratio of the powers in the transduced peak $P(\omega_\mathrm{s})$ and the thermal noise $P(\omega_\mathrm{m})$, where power is the integrated spectral density, is scaled by the number of thermal phonons to find
\begin{equation}
\label{eqn:PSDtophonons}
    \bar{n}_{\mathrm{s}} = \frac{\hbar\omega_\mathrm{m}}{k_\mathrm{B} T} \frac{P(\omega_\mathrm{s})} {P(\omega_\mathrm{m})},
\end{equation}
where $T=295~\mathrm{K}$ is room temperature. To ensure the device is properly thermalized and optical heating is appropriately limited, the thermal noise is in turn calibrated to a temperature-independent electro-optic modulator tone.\cite{MacDonald2016} In Fig.~\ref{fig:3}, the number of actuated phonons are calculated for a range of microwave powers. At low powers, the transduced signal sinks below the thermal noise floor at an average population $(9.0\pm0.4)\times10^{-2}$ actuated phonons. Sufficiently far from this thermal limit, the number of actuated phonons increases linearly with microwave input power. This slope is fit to calculate the single-photon electromechanical coupling rate $g_\mathrm{\upmu}/2\pi = (4.3\pm0.8)~\mathrm{Hz}$. 

The transduction efficiency $\eta_0$ can be calculated from the linearized rotating-frame Heisenberg-Langevin equations of motion for the telecom $a(\omega)$, mechanical $b(\omega)$, and microwave $c(\omega)$ modes:\cite{Jiang2019a}
\begin{align}
i\omega a &= (i\Delta + \kappa/2)a + i g_0 b +\sqrt{\kappa_\mathrm{e}}a_\mathrm{in}(\omega) \label{eqn:a}\\
i\omega b &= (i\omega_\mathrm{m}+\Gamma_\mathrm{m}/2)b + i g_0 a + ig_\mathrm{\upmu}c + \sqrt{\Gamma_\mathrm{m}}b_\mathrm{in}(\omega) \label{eqn:b}\\
i\omega c &= (i\omega_\mathrm{\upmu} + \kappa_\mathrm{\upmu}/2)c +g_\mathrm{\upmu}b+\sqrt{\kappa_\mathrm{\upmu,e}}c_\mathrm{in}(\omega). \label{eqn:c}
\end{align}
In our experiment we set the laser detuning to $\Delta=\omega_\mathrm{c}-\omega_\mathrm{\uell}=0$, and in Fig.~\ref{fig:3} the microwave signal frequency $\omega_\mathrm{s}$ is set to $\omega_\mathrm{\upmu}$. The electromagnetic modes are coupled to input channels $a_\mathrm{in}(\omega)$ and $c_\mathrm{in}(\omega)$, representing telecom laser and electrical signal inputs to the telecom and microwave modes respectively, while $b_\mathrm{in}(\omega)$ represents the influx of thermal phonons into the mechanical mode from the room temperature bath.

Solving Eqns.~(\ref{eqn:a}--\ref{eqn:c}) under these conditions,\cite{Jiang2019a, Aspelmeyer2014, Hauer2015} the equation for microwave-to-telecom efficiency-per-pump-photon for our system is:
\begin{align}
\label{eqn:efficiency}
    \eta_0 = \frac{\kappa_\mathrm{e}}{\kappa}\frac{\kappa_\mathrm{\upmu,e}}{\kappa_\mathrm{\upmu}}
    \frac{4C_\mathrm{\upmu}C_0}{(1+C_\mathrm{\upmu}+C_0)^2 + (2(1+C_\upmu)\omega_\mathrm{m}/\kappa)^2},
\end{align}
where the electromechanical and optomechanical single-photon cooperativities are $C_\mathrm{\upmu} = 4g_\mathrm{\upmu}^2/\kappa_\mathrm{\upmu}\Gamma_\mathrm{m} = (6.2\pm0.2)\times10^{-12}$ and $C_0 = 4g_0^2/\kappa\Gamma_\mathrm{m} = (3.5\pm0.1)\times10^{-4}$ respectively. The single-photon optomechanical coupling rate $g_0/2\pi = (1.3\pm0.3)~\mathrm{MHz}$ was measured using phase calibration.\cite{Ramp2019} Hence, our system efficiency is $\eta_0 = (1.0\pm0.1)\times10^{-15}$.

For high efficiency transduction, the condition $C \approx C_\mathrm{\upmu} \gg 1$ must be achieved, where $C=C_0 \bar{n}_\mathrm{c}$ is the cavity-enhanced optomechanical cooperativity with an average population $\bar{n}_\mathrm{c}$ photons in the telecom mode. The optomechanical cooperativity of this device is predicted to reach $C=3.7$ at millikelvin temperatures.\cite{Ramp2019} However in the present experiment, the telecom photon population of the optomechanical crystal was limited to $\bar{n}_\mathrm{c}\approx400$ to circumvent optical heating, leading to a cooperativity $C=0.14$. The cavity-enhanced transduction efficiency can be calculated by replacing $C_0$ by $C$ in Eqn.~\eqref{eqn:efficiency} to find $\eta_\mathrm{enh} = (3.4\pm0.2)\times10^{-13}$.

Even with the enhancements to optomechanical cooperativity, the efficiency of our system is primarily limited by electromechanical cooperativity, which is in turn limited by the electromechanical coupling rate. The electromechanical coupling is predicated on the overlap of the optomechanical crystal and the microwave electric field.\cite{Zou2016} In our system, this overlap is small due to the large spatial requirement of the dimpled-tapered fiber optical coupling mechanism. Future iterations with permanent fiber coupling \cite{McKenna2019} will allow the gap between the cavity pedestal and lid to be dramatically decreased, thus increasing field overlap and allowing electromechanical coupling to reach rates on the order of $1~\mathrm{kHz}$. Additionally, the electromechanical cooperativity can be further increased by reducing the microwave cavity decay rate and mechanical damping rate. Specifically, 3D superconducting microwave cavities are capable of decay rates below $100~\mathrm{Hz}$, \cite{Reagor2013} and low temperature measurements of the gallium arsenide optomechanical crystals demonstrate a mechanical damping rate of $83~\mathrm{kHz}$.\cite{Ramp2019} With these improvements taken into account, the electromechanical cooperativity could reach values exceeding $1$, which would lead to nearly lossless microwave-to-telecom transduction.

%, permitting coupling rates on the order of megahertz\cite{Zou2016} and efficiency on the order of $1\%$.

\begin{figure}[b]
\includegraphics[width=\columnwidth]{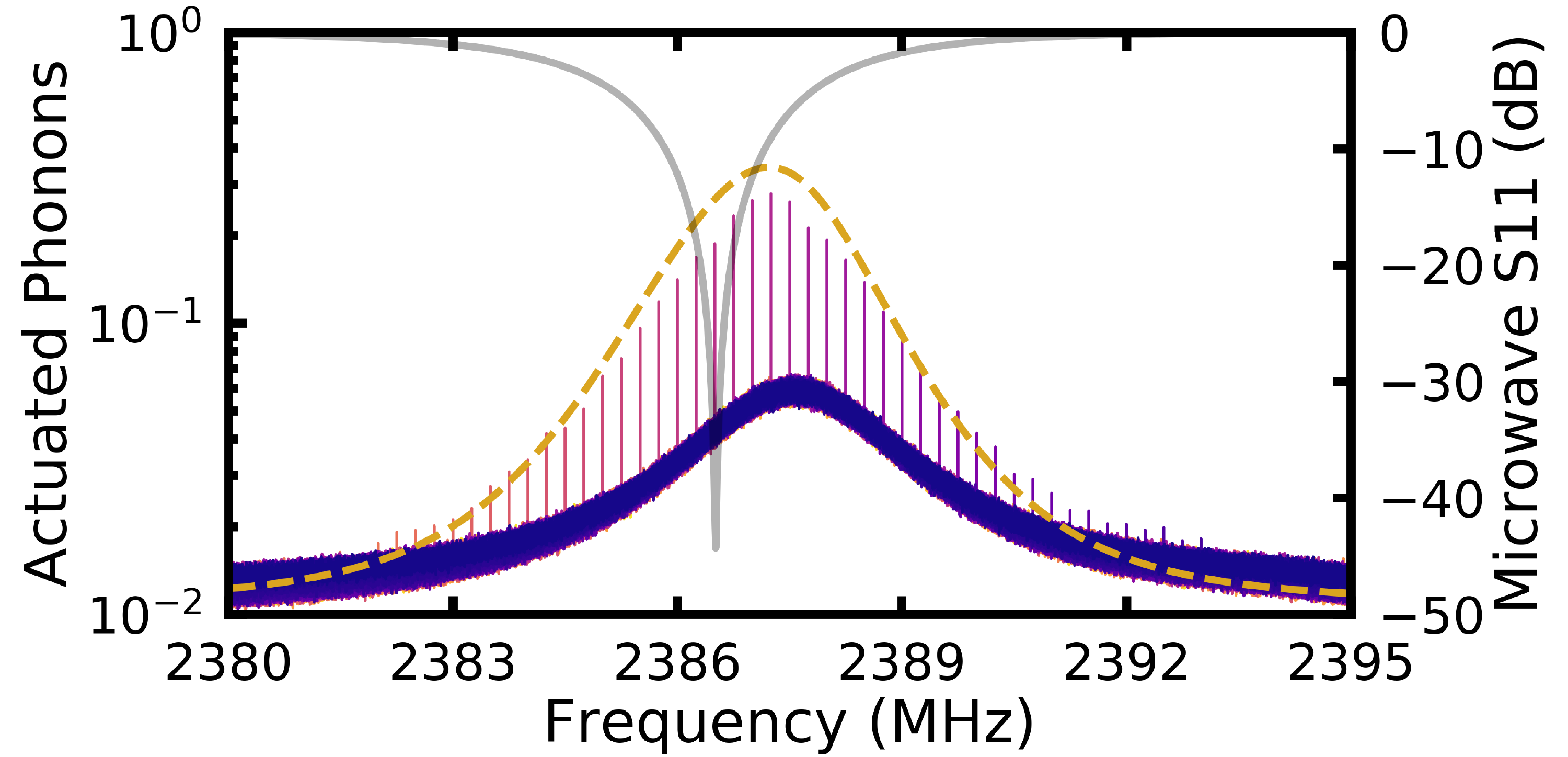}
\caption{\label{fig:4} 
Optical measurements of the transduced $3.2~\mathrm{\upmu W}$ electrical signal as it is swept between $2380~\mathrm{MHz}$ to $2395~\mathrm{MHz}$ in $0.25~\mathrm{MHz}$ steps. The dashed curve represents a calculation of the number of actuated phonons for the whole frequency range, using Eqn.~\eqref{eqn:microwavephonons}  with experimentally measured parameters. 
}
\end{figure}

\begin{figure}[t]
\includegraphics[width=.95\columnwidth]{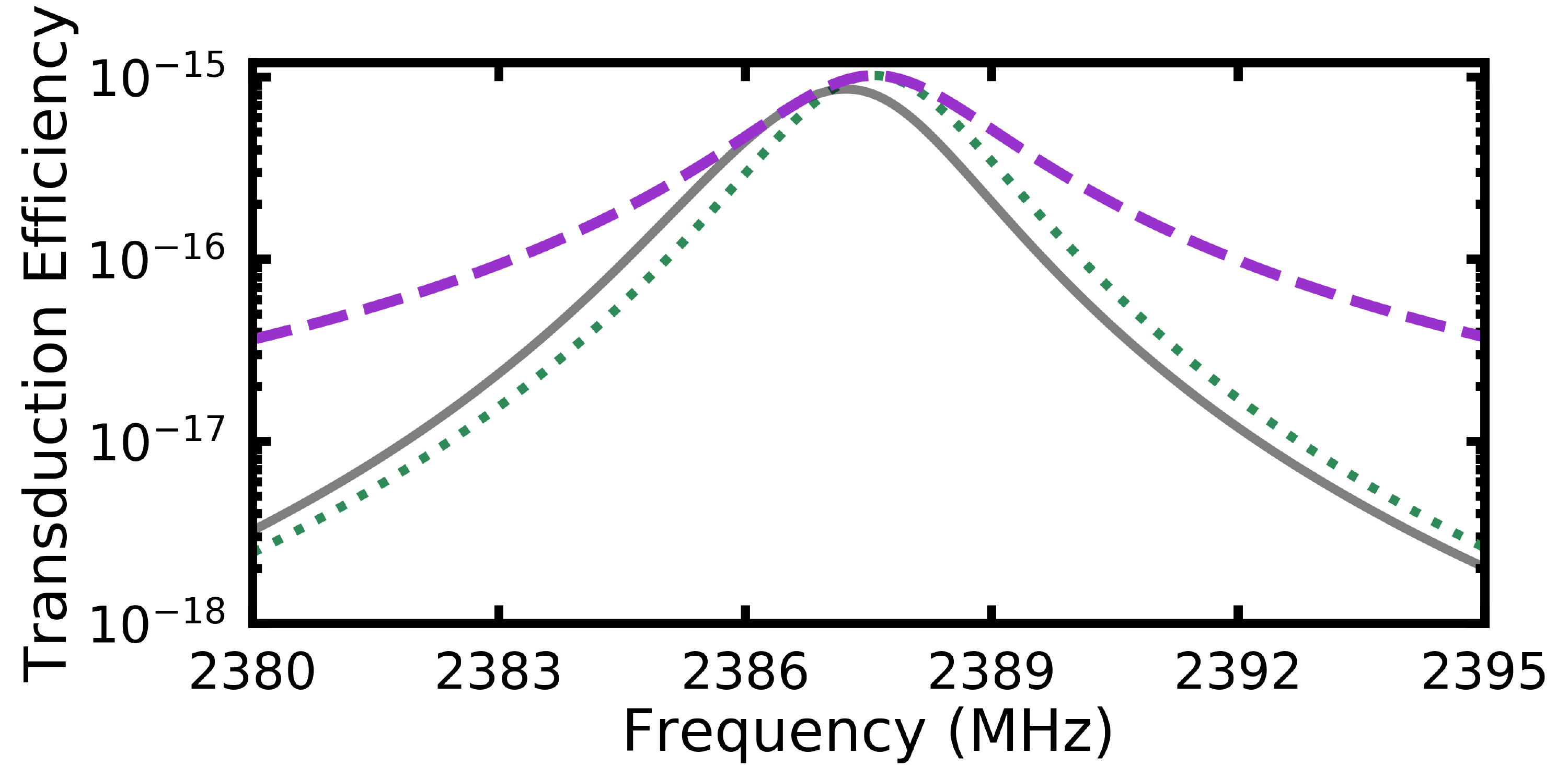}
\caption{\label{fig:5} 
Calculations of the conversion efficiency for the transduction of electrical input signals at various frequencies using the experimental parameters outlined for this experiment (grey solid), the ideal scenario where the detuning between microwave and mechanical resonances is zero (green dotted), and the maximum transduction efficiency attained by using a tunable microwave cavity to select the microwave resonance frequency for maximum transduction efficiency at each input signal frequency (purple dashed).
}
\end{figure}

Another important metric for microwave-to-telecom transduction is the bandwidth over which efficient transduction can be achieved. A large bandwidth increases the utility of the transducer by allowing it to convert a range of microwave frequencies to telecom wavelengths and permits faster transduction operations. \cite{Lauk2019} To measure the bandwidth of transduction, the electrical signal frequency is stepped across the the mechanical and microwave resonances and the mechanical spectrum is measured by optical homodyne at each step. In Fig.~\ref{fig:4} the mechanical spectrum is scaled using Eqn.~\eqref{eqn:PSDtophonons}, such that the amplitude of each transduced peak represents the number of actuated phonons in the mechanical mode. This demonstrates that the maximum number of actuated phonons is achieved for a frequency between the microwave and mechanical peaks at $\omega_\mathrm{max}/2\pi = 2387.25~\mathrm{MHz}$. To confirm this, the equations of motion are used to derive the number of actuated phonons for an electrical input signal with frequency $\omega_\mathrm{s}$ and power $P_\mathrm{s}$,
\begin{align}
    \label{eqn:microwavephonons}
    \bar{n}_\mathrm{s}(\omega_\mathrm{s}) =  \frac{P_\mathrm{s}}{\hbar\omega_\mathrm{s}}
    \frac{g_\mathrm{\upmu}^2\kappa_\mathrm{\upmu,e}} 
    {\left[\left(\omega_\mathrm{m}-\omega_\mathrm{s}\right)^2 + \Gamma_\mathrm{m}^2/4\right]
    \left[\left(\omega_\mathrm{\upmu}-\omega_\mathrm{s}\right)^2 + \kappa_\mathrm{\upmu}^2/4\right]
    }.
\end{align}

In Fig.~\ref{fig:4}, the number of actuated phonons $\bar{n}_\mathrm{s}(\omega_\mathrm{s})$ is plotted across the same region over which the electrical signal is measured to demonstrate the agreement between experiment and theory. From Eqn.~\eqref{eqn:microwavephonons}, a microwave-to-mechanics transduction bandwidth of $2.16~\mathrm{MHz}$ is calculated, which exceeds the predicted bandwidth of similar piezoelectric microwave cavity to telecom transducers. \cite{Wu2019} The large bandwidth is in part due to an efficiency-bandwidth trade-off for transduction: the high decay rate of the microwave cavity and damping rate of the mechanics increases bandwidth, but in turn limits the number of actuated phonons and therefore the efficiency of the transducer. During low temperature experiments both the damping and decay rates will be reduced, which limits the bandwidth of future experiments, but increase transduction efficiency.

In Fig.~\ref{fig:5}, the transduction efficiency is plotted as a function of signal frequency for our experiment where $(\omega_\mathrm{m}-\omega_\mathrm{\upmu})/2\pi \approx 1~\mathrm{MHz}$ and for the ideal transduction scenario $\omega_\mathrm{m}=\omega_\upmu$. Encompassing both the experimental and ideal transduction scenarios is a calculation of efficiency that assumes a tunable 3D microwave cavity\cite{Clark2018} with a resonance frequency set such that the input electrical signal is transduced with maximum efficiency at each considered frequency. In ideal circumstances, where the electrical signal frequency matches the mechanical resonance frequency, the efficiency of transduction is unchanged. For unmatched resonances however, the microwave cavity can be tuned to increase the transduction efficiency. 

The extended range afforded by a tunable cavity has a full-width half-max of $2.91~\mathrm{MHz}$, such that the transduction efficiency of off-resonant signals improves by up to an order of magnitude. Though this is not a true increase in bandwidth, which refers to the maximum frequency spread that can be simultaneously passed by the cavity, it does increase the frequency range over which the transducer can function. For the microwave cavity under consideration in this experiment, the increased transduction range associated with a tunable microwave cavity is modest, but becomes more pronounced when the reduced decay rates of superconducting cavities are taken into account.\cite{Reagor2013} 

In conclusion, we have demonstrated gallium arsenide optomechanical crystals in a 3D microwave cavity as a promising platform for quantum state transduction. The optomechanical crystal is sensitive enough to detect an average occupancy of just $(9.0\pm0.4)\times10^{-2}$ actuated phonons, and is capable of achieving high cooperativity. The piezoelectric coupling $g_\mathrm{\upmu}/2\pi = (4.3\pm0.8)~\mathrm{Hz}$ between the microwave and mechanical modes limits the transduction efficiency to $\eta_0 = (1.0\pm0.1)\times10^{-15}$, but could be improved by reducing the microwave electric field mode volume. Finally, the $2.16~\mathrm{MHz}$ transduction bandwidth of this system allows for a broad range of electrical signals to be transduced. Although this bandwidth will be reduced when the transduction experiment is performed at low temperatures, where the microwave cavity will be superconducting, we have proposed a framework using a tunable microwave cavity to allow for microwave-to-telecom transduction that is efficient, low-loss, broadband, and coherent.

\vspace{10pt}
See the supplementary material for a detailed derivation of the electromechanical coupling rate, bandwidth, and transduction efficiency.
\vspace{5pt}

This work was supported by the University of Alberta; the Natural Sciences and Engineering Research Council, Canada (Grants No. RGPIN-04523-16,No.  DAS-492947-16,  and  No.  CREATE-495446-17); Quantum  Alberta; the Alfred P. Sloan Foundation;  and  the  Canada  Foundation  for Innovation.

\section{Supplement: Calculation of electromechanical coupling and transduction efficiency}

The equations of motion, Eqns.~(2--4) in the main text, are derived from the linearized optomechanical Hamiltonian in a frame rotating at the laser frequency with an added term to include the piezo-mechanical coupling,\cite{Carvalho2019}
\begin{align}
    \label{eq:Hamiltonian}
     \mathcal{H} &= \hbar\Delta a^\dagger a + \hbar\omega_\mathrm{m}b^\dagger b + \hbar\omega_\mathrm{\upmu}c^\dagger c\nonumber\\
     &+
     \hbar\left[g_0 \sqrt{\bar{n}_\mathrm{c}} (a^\dagger + a) + g_\mathrm{\upmu}(c^\dagger+c)\right](b^\dagger+b),
\end{align}
where in what follows we have set the number of telecom cavity photons to $n_\mathrm{c}=1$. The cavity enhanced equations can be recovered in the following work by replacing $g_0 \rightarrow g_0 \sqrt{\bar{n}_\mathrm{c}}$.
%where $g$ is the optomechanical coupling enhanced by the number of photons in the cavity $g=g_0\sqrt{n_\mathrm{cav}}$. 

To determine the electromechanical coupling rate and transduction efficiency, we begin with the equations of motion solved for their respective operators,\cite{Jiang2019a,Aspelmeyer2014}
\begin{align}
    \label{eqn:asolved}
    a(\omega) &= -
    \frac{ig_0 b(\omega) + \sqrt{\kappa_\mathrm{e}}a_\mathrm{in}(\omega)}
    {\chi^{\shortminus 1}_a(\omega)},
    \\
    \label{eqn:bsolved}
    b(\omega) &= -
    \frac{ig_0 a(\omega) + ig_\mathrm{\upmu} c(\omega) + \sqrt{\Gamma_\mathrm{m}}b_\mathrm{in}(\omega)} 
    {\chi^{\shortminus 1}_b(\omega)},%{i\left(\omega_\mathrm{m}-\omega\right) + \Gamma_\mathrm{m}/2} 
    %+\frac{\sqrt{\Gamma_\mathrm{m}}b_\mathrm{in}(\omega)} 
    %{\left(\omega_\mathrm{m}-\omega\right) + \Gamma_\mathrm{m}/2}
    \\
    \label{eqn:csolved}
    c(\omega) &= -
    \frac{ig_\mathrm{\upmu}b(\omega) + \sqrt{\kappa_\mathrm{\upmu,e}}c_\mathrm{in}(\omega)}
    {\chi^{\shortminus 1}_c(\omega)},%{i\left(\omega_\mathrm{\upmu} - \omega\right) + \kappa_\mathrm{\upmu}/2}.
\end{align}
where counter-rotating terms have been omitted for simplicity, and the inverse susceptibilities are defined as 
\begin{align}
    \chi^{\shortminus 1}_a(\omega) &= i\left(\Delta-\omega\right) + \kappa/2, \\
    \chi^{\shortminus 1}_b(\omega) &= i\left(\omega_\mathrm{m}-\omega\right) + \Gamma_\mathrm{m}/2,\\
    \chi^{\shortminus 1}_c(\omega) &= i\left(\omega_\mathrm{\upmu}-\omega\right) + \kappa_\mathrm{\upmu}/2.
\end{align}

We now proceed with the calculation of the electromechanical coupling rate. Since we have considered the system to be in a state where the telecom mode of the optomechanical crystal is sparsely populated, the term $g_0 a(\omega)$ is negligible. Moreover, we assume the microwave mode of the 3D cavity is predominantly populated with microwave input photons, such that $g_\mathrm{\upmu}b(\omega) \ll \sqrt{\kappa_\mathrm{\upmu,e}}c_\mathrm{in}(\omega)$. With these simplifications made, we focus on the mechanics. Eqn.~S3 becomes
\begin{align}
    \label{eqn:bsolvedfurther}
    b(\omega) = 
    \frac{ig_\mathrm{\upmu}\sqrt{\kappa_\mathrm{\upmu,e}}c_\mathrm{in}(\omega)} 
    {
    \chi^{\shortminus 1}_b(\omega)%\left[i\left(\omega_\mathrm{m}-\omega\right) + \Gamma_\mathrm{m}/2\right] 
    \chi^{\shortminus 1}_c(\omega)%\left[i\left(\omega_\mathrm{\upmu}-\omega\right) + \kappa_\mathrm{\upmu}/2\right] 
    }%\nonumber\\
    -
    \frac{\sqrt{\Gamma_\mathrm{m}}b_\mathrm{in}(\omega)} 
    {
    \chi^{\shortminus 1}_b(\omega)%{i\left(\omega_\mathrm{m}-\omega\right) + \Gamma_\mathrm{m}/2}.
    }.
\end{align}

The power spectral density function associated with the negative-frequency sideband of the microwave-driven mechanical mode is given by\cite{Hauer2015}
\begin{equation}
    S_{b^\dagger b}(\omega) = \frac{1}{2\pi} \int \braket{ b^\dagger (\omega) b (\omega^\prime)} d\omega^\prime,
\end{equation}
which can be solved using the correlators for the thermal and microwave input
\begin{align}
    \braket{b^\dagger_\mathrm{in}(\omega)b_\mathrm{in}(\omega^\prime)} &= 2\pi \bar{n}_\mathrm{b}(\omega_\mathrm{m})\delta(\omega+\omega^\prime),\\
     \braket{c^\dagger_\mathrm{in}(\omega)c_\mathrm{in}(\omega^\prime)} &= (2\pi)^2 \frac{P_\mathrm{s}}{\hbar\omega_\mathrm{s}}\delta(\omega + \omega_\mathrm{s}) \delta(\omega^\prime - \omega_\mathrm{s}). 
\end{align} 
Where $\bar{n}_\mathrm{b}(\omega_\mathrm{m}) \approx k_b T/\hbar\omega_m$ is the mean thermal occupancy of thermal bath at the mechanical frequency, and $P_\mathrm{s}$ is the microwave power at frequency $\omega_\mathrm{s}$ incident on the microwave cavity. Here, the thermal photon population of the microwave cavity is negligible compared to the number of photons created by the microwave drive.

We solve the terms in Eqn.~S8 individually, noting that correlations between the microwave input and thermal noise are zero. The total spectra is then the summation of two peaks, the first resulting from the actuated phonons from the microwave drive at frequency $\omega_\mathrm{s}$:
\begin{align}
    \label{eqn:mwspectra}
    S_{b^\dagger b}^{\mathrm{s}}(\omega) &= \frac{P_\mathrm{s}}{\hbar\omega_\mathrm{s}}
    \frac{%numerator
    2\pi g_\mathrm{\upmu}^2\kappa_\mathrm{\upmu,e} \delta(\omega+\omega_\mathrm{s})
    } 
    {%denominator
    [\chi^{\shortminus 1}_b(-\omega)]^* [\chi^{\shortminus 1}_c(-\omega)]^* \chi^{\shortminus 1}_b(\omega_\mathrm{s})\chi^{\shortminus 1}_c(\omega_\mathrm{s})
%    \left[  \left(\omega_\mathrm{m}-\omega\right)^2 + \Gamma_\mathrm{m}^2/4\right]
%    \left[\left(\omega_\mathrm{\upmu}-\omega\right)^2 + \kappa_\mathrm{\upmu}^2/4\right]
    },  
\end{align}
and the second a result of the thermal background:
\begin{align}
    \label{eqn:thspectra}
    S_{b^\dagger b}^{\mathrm{th}}(\omega) &=
    \frac{\Gamma_\mathrm{m}\bar{n}_\mathrm{b}(\omega_\mathrm{m})} {\left(\omega_\mathrm{m}+\omega\right)^2 + \Gamma_\mathrm{m}^2/4}.
\end{align}

The average number of actuated phonons in the mechanical mode can be calculated from $\bar{n}_\mathrm{s}=(2\pi)^{-1}\int S_{b^\dagger b}^\mathrm{s}(\omega) d\omega $. For a narrow-frequency signal input $\omega_\mathrm{s}$, the actuated phonon occupancy is

\begin{align}
    \label{eqn:app:microwavephonons}
    \bar{n}_\mathrm{s}\left(\omega_\mathrm{s}\right) =  \frac{P_\mathrm{s}}{\hbar\omega_\mathrm{s}}
    \frac{g_\mathrm{\upmu}^2\kappa_\mathrm{\upmu,e}} 
    {\left[\left(\omega_\mathrm{m}-\omega_\mathrm{s}\right)^2 + \Gamma_\mathrm{m}^2/4\right]
    \left[\left(\omega_\mathrm{\upmu}-\omega_\mathrm{s}\right)^2 + \kappa_\mathrm{\upmu}^2/4\right]
    }.
\end{align}
If $\omega_\mathrm{s}$ is swept across the microwave and mechanical resonances, then Eqn.~S14 forms the envelope plotted in Fig.~4. Similarly, if $\omega_\mathrm{s}$ is fixed at $\omega_\mathrm{\upmu} \approx \omega_\mathrm{m}$, then Eqn.~S14 simplifies to
\begin{equation}
    \label{eqn:coh_phonons}
    \bar{n}_\mathrm{s} = \left( 
    \frac{16 g_\mathrm{\upmu}^2 \kappa_\mathrm{\upmu,e}}
    {\Gamma_\mathrm{m}^2 \kappa_\mathrm{\upmu}^2 \hbar\omega_\mathrm{s}}
    \right) P_\mathrm{s},
\end{equation}
which is the linear equation relating the number of microwave-induced phonons to input microwave power. Using the slope in Fig.~3, we calculate the electromechanical coupling $g_\upmu/2\pi = (4.3\pm0.8)~\mathrm{Hz}$.

The efficiency is also derived from Eqns.~(\ref{eqn:asolved}-\ref{eqn:csolved}). Here we refer the reader to Ref.~\onlinecite{Jiang2019a}, where the general efficiency for a microwave-mechanical-telecom system is derived to be 
\begin{align}
    \eta = \left|
    \frac{g_0 g_\upmu \sqrt{\kappa_\mathrm{e}}  \sqrt{\kappa_\mathrm{\upmu,e}}}
    {
    (\chi_a(\omega)\chi_b(\omega)\chi_c(\omega))^{-1}
    +\chi^{\shortminus 1}_c(\omega) g_0^2
    +\chi^{\shortminus 1}_a(\omega) g_\mathrm{\upmu}^2
    }
    \right|^2.
\end{align}
In our case, we solve with laser detuning $\Delta = 0$, and assume a near-resonant coupling condition such that $\omega = \omega_\mathrm{m} = \omega_\mathrm{\upmu} $ to obtain Eqn.~5 from the main text. It is worth pointing out that Eqn.~5 differs from the typical red-detuned scenario (where $\Delta = \omega_\mathrm{m}$) by the addition of $(2(1+C_\mu)\omega_\mathrm{m}/\kappa)^2$ in the denominator. Our choice of measuring the telecom optical cavity on-resonance does therefore decrease the efficiency, but only by a small amount: $\eta_{\Delta=0}/\eta_{\Delta=\omega_\mathrm{m}} \approx 0.66$. The advantage of this choice is that there are no optomechanical back action effects which alter the total mechanical damping rate, allowing us to make the approximation $\Gamma_\mathrm{tot} = \Gamma_\mathrm{m}$ and simplify the mathematical treatment of the system.

\end{document}